\documentstyle[12pt, aasms4, psfig]{article}

\begin{document}

\title{Contemporaneous UV and Optical Observations of Direct and Raman Scattered
O VI Lines in Symbiotic Stars}

\author{Jennifer J. Birriel}
\affil{Department of Physics \& Astronomy, University of Pittsburgh, Pittsburgh
PA 15260; jennifer@phyast.pitt.edu}

\author{Brian R. Espey} \affil{Space Telescope Science Institute, 3700
San Martin Drive, Baltimore MD, 21218; espey@stsci.edu}

\author{Regina E. Schulte-Ladbeck} \affil{Department of Physics \&
Astronomy, University of Pittsburgh, Pittsburgh PA 15260;
rsl@phyast.pitt.edu}

\begin{abstract}

Symbiotic stars are binary systems consisting of a hot star, typically
a white dwarf, and a cool giant companion.  The wind from the cool
star is ionized by the radiation from the hot star, resulting in the
characteristic combination of sharp nebular emission lines and stellar
molecular absorption bands in the optical spectrum.  Most of the
emission lines are readily identifiable with common ions. However, two
strong, broad emission lines at $\lambda\lambda$6825 and 7082 defied
identification with known atoms and ions.  In 1989, Schmid made the
case that these long unidentified emission lines resulted from the
Raman scattering of the O VI resonance photons at
$\lambda\lambda$1032, 1038 by neutral hydrogen.

We present contemporaneous far-UV and optical observations of direct
and Raman scattered O~VI lines for nine symbiotic stars obtained with
the Hopkins Ultraviolet Telescope (Astro-2) and various ground-based
optical telescopes.  The O~VI emission lines are present in every
instance in which the $\lambda$$\lambda$6825, 7082 lines are present,
in support of the Schmid Raman scattering model.  We calculate the
scattering efficiencies and discuss the results in terms of the Raman
scattering model.  Additionally, we measure the flux of the Fe~II
fluorescence line at $\lambda$1776, which is excited by the O VI line
at $\lambda$ 1032, and calculate the first estimates of the conversion
efficiencies for this process.

\end{abstract}


\section{INTRODUCTION}

Roughly half of all known symbiotic stars exhibit two strong, broad
emission lines at 6825 and 7082\AA\ (Allen 1980).  These emission
features are much broader than other emission features ($\sim$ 20\AA)
and are often among the ten strongest lines in the optical spectrum.
These lines are only observed in symbiotic stars and, of those, only
in systems that exhibit other high excitation features such as [Ne V]
and [Fe VII].  For years the identity of these lines remained
uncertain and they simply became known as the``unidentified''
lines.

Schmid (1989) proposed that the long unidentified emission lines in
the optical spectra of some symbiotic stars at 6825, 7082\AA\ result
from the Raman scattering of O~VI resonance photons at 1032, 1038\AA\
by neutral hydrogen.  The Raman process involves the inelastic
scattering of O~VI photons produced near the hot component by neutral
hydrogen atoms near the cool component of the binary system.  An O~VI
photon excites a neutral hydrogen atom from the ground state to a
virtual state between the n=2 and n=3 levels, followed by an immediate
emission of a photon, which leaves the hydrogen atom in the excited
2s$^2$ S-state.  From energy conservation, the frequency of the
scattered photon is the difference between the energy of the virtual
and final states of the atom or, in wavelength terms:
$\lambda_{\mathrm{O~VI}}^{-1} - \lambda_{\mathrm{Ly\alpha}}^{-1}=
\lambda_{\mathrm{Raman}}^{-1}$.  This change in energy by $\sim$6.7
also results in a corresponding increase in the linewidth of the
scattered lines, producing two broad lines in the red region of the
spectrum (Nussbaumer, Schmid, \& Vogel 1989).

Raman scattering is a dipole process which produces polarization.  The
Raman lines at 6825 and 7082~\AA\ are generally observed to be
strongly polarized, typically being 6\% (Schmid 1996 and references
therein).  This strong polarization results because the scattering
geometry in symbiotic systems is anisotroptic.  The O~VI far-UV
photons are produced near the hot star and Raman scattered near the
cool star. Additionally, all the Raman photons are produced by a
scattering process, so the polarization is not diluted by light coming
from an unscattered source.  The strong polarization of the emission
lines at 6825 and 7082~\AA\ has been observed and extensively
investigated in a number of symbiotics by several authors (Schmid \&
Schild 1994, 1997a,b, 2000; Harries \& Howarth 1996a,b) and modeled in
a number of papers (Schmid 1996, Harries \& Howarth (1997), Lee \& Lee
1997a,b).

Due to the small cross-section of Raman scattering, the main
prerequisite for the production of the 6825, 7082\AA\ emission is the
presence of strong O~VI emission.  Indirect evidence of O~VI emission
in symbiotic systems includes the presence of O~V~$\lambda$~1371
emission that arises in part from dielectronic recombination from O~VI
(Hayes \& Nussbaumer 1986), and the presence of Fe~II transitions which
are excited by O~VI photons (Johansson 1988; Fiebelman, Bruhweiler, \&
Johansson 1991).  Far-UV observations of a few symbiotic stars using
{\it Voyager} found strong evidence for Ly$\beta$/O~VI emission ( Li
\& Leahy 1997), however, the low resolution, $\sim$20~\AA, made it
difficult to determine the exact amount of O~VI and Ly$\beta$ present
in each object.  In 1993 the ORFEUS-I mission made unambiguous
detections of O~VI in a small number of symbiotic stars (Schmid et
al. 1999) and, during the flight of the Astro-2 in 1995, Espey et
al. (1995) used far-UV data from the Hopkins Ultraviolet Telescope or
HUT and optical data from the Anglo-Austrialian Telescope (AAT) to
provide the first simultaneous observations of both the O~VI 1032,
1038 \AA\ and scattered 6825, 7082 \AA\ lines in the
symbiotic star RR~Tel.  More recently, Birriel et al.  (1998) used
far-UV data from the HUT and near-simultaneous optical data to
determine the Raman scattering efficiency in the symbiotic star Z~And.
Schmid et al.\ (1999) presented far-UV observations from ORFEUS-I and
II showing strong O VI emission for six symbiotic stars and used
available, though not necessarily contemporaneous, optical observations to
determine scattering efficiencies for the Raman process.

In this paper we examine contemporaneous ultraviolet and optical
observations for nine symbiotic stars taken with the HUT (Astro-2) and
various ground-based optical telescopes.  In support of the Schmid
Raman scattering hypothesis, we find no case in which the 6825, 7082
\AA\ lines appear in the optical without the O VI 1032, 1038\AA\
emission simultaneously occuring in the far-UV spectral region.  We
calculate scattering efficiencies where possible, compare these to the
results of Schmid et al.  (1998) and discuss possible scattering
geometries.  In addition, we make the first determinations of the
efficiency for the Bowen fluorescence process in which a O~VI
1032~\AA\ photon excites the Fe~II 1776.5~\AA\ line (actually an
unresolvable blend of the 1776.660~\AA\ and 1776.418~\AA\ lines).

\section{OBSERVATIONS}

Far-UV data from the HUT were obtained during the Astro-2 mission in
March 1995 (see Table 1).  HUT data cover the wavelength range of
825-1850\AA\ with a nominal resolution of $\approx 3$~\AA.  A detailed
description of the HUT and its calibration and performance during the
Astro-2 mission can be found in Davidsen et al.\ (1992) and Kruk et
al.\ (1995), respectively.  Kruk et al. (1999) provide a description
of the acquisition of the ultraviolet spectrum of each object.  Espey
et al. (1995) describe the data processing for HUT spectra.

The HUT spectra used in this study are listed in Table 1a.  We list
relevant details for each observation including acquisition date,
exposure time, slit diameter, and photometric corrections.  The HUT
symbiotic spectra were obtained during orbital day and hence contain
strong airglow emission lines.  In some observations, pointing errors
resulted in flux loss.  The HUT team determined photometric correction
factors to acount for these losses and the final calibrated  spectra
for most symbiotic stars have flux accuracies of $\sim$5\% (Kruk et
al. 1999).   During the observation of CH Cyg the measured flux
varied gradually, rising and falling by almost 50\%.  The pointing was
fairly stable and the flux variations were not correlated with any
changes in pointing, thus the photometric correction listed in Table 1
is probably not meaningful.  For most objects, the signal-to-noise in
the continuum region near the O~VI emission is $\sim$10 per pixel
($\sim$0.51\AA).

Many of the optical observations for this study were taken around the
time of the Astro-2 mission by Miko{\l}ajewska \& Kenyon (see Table
1b).  The data were obtained using the FAST spectrograph mounted on
the 1.5-m telescope at the Fred L.  Whipple Observatory on Mount
Hopkins, AZ.  These data cover a spectral range 3800-7500\AA\ at a
resolution of $\sim$3~\AA.  The symbiotic spectra were calibrated
using FAST spectra of several Hayes \& Latham (1975) flux standards.

Some of the optical observations on 3 Jan 1995 and 3 April 1995 were
affected by the presence of thin clouds which resulted in inaccurate
spectrophotometry.  We used a synthetic photometry program written by
one of us (BRE) to estimate the scalar correction required to bring
the derived photometry into agreement with the broad-band photometry
of Hric et al.\ (1996a; 1996b).  The applied photometric scale factor
is indicated in Table 1b.   For Z And and CH Cyg, both of which have
very good photometric coverage, agreement between observed and
synthetic photometry (using B, V, and sometimes R magnitudes) is
better than 10\%, giving us confidence in our synthetic photometry
estimates.  For EG And, the photometric coverage is sparse, but
brackets the time of observation.  Amateur observations by the
Association Fran\c{c}ais d'Etoiles Variables (available on the web at
ftp://cdsarc.u-strasbg.fr/pub/afoev/) were used to estimate the
variation between the photometric data.  For AG Dra and AX Per, on the
other hand, very good coverage exists, with photometric observations
within a week or two of the ground-based data.  Simple linear
interpolation was used to estimate the fluxes at the time the
spectroscopic data were taken.  In view of the calibration required
for these data, flux errors of $\approx 20$\% are probably a
reasonable assumption.  We compared our recalibrated spectra with
others obtained close in time (or phase) to these observations and
found that continuum and emission line fluxes agree to within
$\sim$20\%.
  
Contemporaneous optical observations are not available for HM~Sge.
Schmid \& Schild obtained an optical spectrum of HM~Sge on the night
of 15 July 1995 with the 4.2-m telescope of the Royal Greenwich
Observatory on La Palma, Canary Islands.  The spectral resolution was
0.65~\AA.  The emission at $\lambda\lambda$ 6825, 7082 is very weak,
and the data are not fluxed, thus we cannot derive an estimate for the
scattering efficiency.  Lower resolution spectra obtained using the
WIYN telescope nearly a year later also show a very weak feature in
the region of $\lambda$ 6825~\AA\ while the emission feature at
$\lambda$7082~\AA\ is not visible.

\section{ANALYSIS}

The data were measured using the {\tt SPECFIT} $\chi^2$-minimization
routine (Kriss 1994) running under the IRAF\footnote{The Image
Reduction and Analysis Facility (IRAF) is distributed by the National
Optical Astronomy Observatories, which is operated by the Association
of Universities for Research in Astronomy, Inc.\ (AURA) under
cooperative agreement with the National Science Foundation.}  data
reduction package.  SPECFIT has the advantage that a variety of
functional forms can be used to describe the continuum, emission
lines, and absorption lines.  Additionally, user-defined functions can
be used to model the absorption features and airglow emission lines.
Observed fluxes for the O~VI lines were measured assuming a simple
linear fit to the continuum and Gaussian profiles for the 1032,
1038\AA\ lines.  At the modest resolution of the HUT, ($\sim$3 \AA\ at
the O~VI doublet), the O~VI lines are partially blended and the
linewidths unresolved (see Figure 1).  To measure the O~VI components,
we tied the ratio of the emission line wavelengths at the theoretical
value (Morton 1991) and the width of the weaker line was tied to that of
the stronger line.

Observed fluxes for the optical lines at 6825, 7082~\AA\ were measured
using a similar approach. However, the fit to the continuum depended
on the spectral type of the cool red giant star. For objects with a
K-type giant star, such as AG~Dra, V1016~Cyg, HM~Sge, and RR~Tel, the
continuum was modelled with a simple linear fit in the region near the
emission features.  The 6825, 7082\AA\ emission lines were fitted with
Gaussian profiles.  In the case of Z~And, EG~And, BF~Cyg, CH~Cyg, and
AX~Per the optical continuum is that of a M-type giant star.  Kenyon
\& Fernandez-Castro (1987) examined the depth of TiO and VO absorption
bands in the red spectra of symbiotics and concluded that most
symbiotic stars contain normal red giants.  We used SPECFIT to find
the best-fit spectra from the observed sample of Fluks et al.\ (1994)
to the optical continua of the Z~And, EG~And, BF~Cyg, CH~Cyg, and
AX~Per.  Table 5 shows our derived optical spectral types for these
objects and compares them to the results of Kenyon \&
Fernandez-Castro.  The spectral type of the red giant probably depends
on illumination effects and so we give the orbital phase of the
optical observations used here; Kenyon \& Fernandez-Castro give
spectral types based on spectra obtained near the photometric minimum
for each object.  The 6825, 7082\AA\ lines were represented by
Gaussians which were tied together by width and by their rest
wavelength ratio.

Table~2 gives the observed fluxes of both the direct and Raman
scattered O~VI lines.  In general the UV line fluxes are good to 10\%
or better and the optical line fluxes to 20\% or better.  The HUT
far-UV spectra and the optical spectra are presented in Figures 1--3.

\subsection{Interstellar H$_2$ Absorption in the O~VI Line Region}

For comparison with other authors, we list in Table 2 the observed
fluxes of the O~VI lines as measured relative to a local linear
continuum in the HUT spectra.  We expect the ratio of the O~VI
1032\AA\ and the 1038\AA\ lines to be somewhere between the optically
thin (2:1) and thick (1:1) values, but a quick glance at Table 2 shows
that the observed ratios for Z~And, V1016~Cyg, and AG~Dra exceed the
optically thin value of 2:1.  Schmid et al. (1999) also find that the
observed O~VI flux ratios exceed the optically thin ratio; in
addition, their observed ratios are rather different than those found
in this study.  The discrepancies between the HUT and ORFEUS data may
due to variations in the O~VI line fluxes; this is discussed further
in \S 3.4.

We attribute these anomalously high (f(1032)/f(1038) $>$2.0) O~VI flux
ratios for Z~And, V1016~Cyg, and AG~Dra, to attenuation by
interstellar absorption.  The far-ultraviolet wavelength region where
the O~VI emission lies is subject to significant absorption due to
interstellar atoms and molecules, in particular H$_2$.  The 1032~\AA\
line is relatively uncontaminated, whereas the line at 1038~\AA\ sits
in a strong absorption band (Morton 1975).  In the HUT data for Z~And
and V1016~Cyg we see absorption features attributable to H$_2$ at
wavelengths below 1130~\AA, though these features are not quite as
evident in the spectrum of AG~Dra.  The spectra of CH~Cyg and RR~Tel
show only weak, if any, absorption as these objects are located high
above the Galactic plane.

Determination of scattering efficiencies for both the 1032~\AA\ and
1038~\AA\ lines requires that we know the actual fluxes emitted within
the region near the hot star, before any extinction and/or absorption
due to interstellar dust and gas.  Hence we must account for
interstellar H~I and H$_2$ absorption in order to determine the
scattering efficiencies for the Raman process in Z~And, V1016~Cyg, and
AG~Dra.  A fit to the UV continuum of each object allows us to
estimate the amount of absorption due to these species.  We can then
perform a simple calculation to correct the emission line fluxes for
H$_2$ absorption.

\subsection{UV Continuum Fits} 

Two sources contribute to the ultraviolet continuum in the HUT
spectral region: the hot star and the nebula.  We use a simple
blackbody spectrum for the hot star and model the nebular contribution
assuming that the dominant source is H~I Balmer recombination
radiation.  The resultant continuum is attenuated using the extinction
curve of Cardelli, Clayton and Mathis (1989, hereafter CCM).  For
continuum fitting we choose regions free of both emission and
absorption line features.

The free parameters for the fitting procedure are the blackbody
temperature, the flux emitted by the hot star at 5500~\AA, the
electron temperature of the nebula, the flux at the Balmer edge (3646
\AA), and the extinction.  The flux emitted by the hot star is a
function of the hot star temperature, the stellar radius, and the
star's distance; likewise, the flux emitted by the nebula is a
function of the nebular electron temperature, nebular size, etc.  In
order to reduce the number of free parameters, and thus the complexity
of the fitting procedure, we fix the values of the hot star
temperature and the nebular electron temperature.  We applied the
modified Zanstra method of M\"urset et al. (1991) together with the
equivalent width of the He~II 1640~\AA\ line to determine the hot star
temperature.  In Table 3 we compare our derived values with those of
M\"urset et al.  For AG~Dra and CH~Cyg, our derived temperatures are
much lower than found by M\"urset et al.  We use their value because
our observations of AG~Dra and CH~Cyg were obtained during outburst
(Tomova \& Tomov 1999, Gonz\'{a}lez-Riestra et al. 1999, Greiner et
al. 1997; Espey et al. 1995b, Munari et al. 1996). The modified
Zanstra method is not valid during outburst as the enhancement in the
continuum is greater than that in the He~II line, and the continuum
shape changes rapidly during outburst; this problem is discussed in
greater detail in M\"{u}rset et al.  (1991).  Additionally,
Miko{\l}ajewska et al.  (1995) point out that the intensities of
various high ionization lines such as He~II and N~V require a hot star
temperature in excess of 10$^5$~K.  In modeling the nebular
recombination continuum radiation, we must specify the electron
temperature, typically $\sim~1.5 ~\times~10^4$~K; here we chose values
found in the literature (Miko{\l}ajewska \& Kenyon 1996,
Miko{\l}ajewska et al. 1995, and Nussbaumer \& Schild 1981).

The spectrum below 1130~\AA\ is affected by the presence of
interstellar H~I and H$_2$. We fit the absorption lines by scaling a
template consisting of optical depth {\it vs.\/} wavelength.  Our
template is derived from recent theoretical data of H$_2$\ rotational
transitions of Abgrall et al. (1993a, b), broadened to a FWHM of
5~${\rm km\ s^{-1}}$ which is typical of cool interstellar gas (Morton 1975,
van Dishoeck \& Black 1986 and references therein).  The relative
scaling of the optical depths comes from the observed interstellar
H$_2$ column densities obtained from ORFEUS-II observations of
Galactic stars (Dixon et al. 1998).  The resulting theoretical model
is convolved with the instrumental resolution (Kruk et al.\ 1998),
before fitting to spectral regions affected solely by H$_2$. Once the
overall H$_2$ column density is obtained, its contribution is fixed,
and absorption due to H~I is determined by fitting the spectral
regions affected by its influence alone.

We successfully fit the HUT UV continua of Z~And, EG~And, BF~Cyg,
V1016~Cyg, AG~Dra, and AX~Per and the derived extinction values for
these objects are found in Table 4.  The fit to the UV-spectra in the
O~VI line region is displayed in figures 1-3 as a dotted line
overplotted on the HUT data.  We did not fit the HUT UV continuum of
RR~Tel as described above; its observed line ratio does not indicate
significant H$_2$ absorption and there are no observable H$_2$
absorption bands in its far-UV spectral region.  We were unable to
obtain continuum fits for the HUT spectra of HM~Sge and CH~Cyg.  The
continuum of HM~Sge is too flat and too weak in the far-UV region to
allow for fitting.  Above 1200\AA\ the HUT UV spectrum of CH~Cyg shows
P-Cygni absorption features and a continuum shape typical for late
A-type supergiants and we therefore do not fit this object's continuum
with our hot star-nebular continuum model.

We are mostly concerned with those objects which exhibit the Raman
scattering effect, i.e. those objects with both the O~VI in the far-UV
and the emission at 6825\AA\ and 7082\AA.  Our best fits to the
continua of Z~And, V1016~Cyg, and AG~Dra are found in Table 6; here we
list the derived parameters of extinction, H~I column density, and
H$_2$ column density and the values of the adopted hot star
temperature and nebular electron temperature.  The continuum fits for
the region below 1130\AA\ are illustrated in Figure 4; these fits
consist of a blackbody spectrum reddened by the CCM (1989) extinction
curve and include interstellar H~I and H$_2$ absorptions and
terrestrial N~I, O~I, and Ly $\beta$ airglow.

\subsection{O~VI Emission Lines and H$_2$ Absorption Corrections}

Our method of correcting for unresolved molecular hydrogen absorption
will not work for emission lines. In the emission line
case, the discrete nature of the emission lines, and their finite width
relative to the absorption, precludes direct estimation of the
contamination from their unresolved profiles.  Correcting the O~VI
emission lines for molecular absorption requires a number of steps and
follows the approach discussed in Birriel, Espey, and Schulte-Ladbeck
(1998).  The absorbed O~VI line strength is measured using the
absorbed continuum discussed in Section 3.3, next a correction factor
is determined, then the final line flux is obtained by correcting the
flux for the presence of molecular hydrogen lines.

The H$_2$ template is generated at 0.01 \AA\ resolution with column
density set to the values used for the continuum fit (Table 6).  The
emission lines are modeled as Gaussians with intrinsic widths of
$\sim$70~km/s based on the prelimary estimates from Orfeus-I and -II
data (Schmid et al.  1999).  It is most likely that the relative
velocity of any molecular cloud and the emitting symbiotic will be
non-zero; since the cloud lies somewhere between the earth and the
symbiotic, we expect that its velocity will be somewhere between that
of the sun and the symbiotic star.  We assume the relative velocity of
the absorption and emission components to be about half the
heliocentric radial velocity of the symbiotic system.  Heliocentric
radial velocities for Z~And, V1016~Cyg, and AG~Dra are taken from the
high-resolution echelle spectra of Ivison, Bode, \& Meaburn (1994).
The amount of absorption of each far-UV emission line is estimated by
integrating the product of the emission and absorption line profiles
across each emission line.  In general, the $\lambda$1032 line suffers
only slight absorption ($\sim$10\%) while the flux in $\lambda$1038 line is
significantly reduced.

Table 7 lists the transmission factors which have also been
incorporated into the corrected O~VI flux values shown in Table 8.
The corrections for H$_2$ made in this study assume a single cloud
along the line of sight and thus do not account for multiple cloud
absorptions or circumstellar absorption.  The revised flux ratios for
Z~And, V1016~Cyg, and AG~Dra are closer to the optically thin ratio
(2:1), see Table 8.  We point out that under the Raman scattering
model of Schmid (1996) the in-situ relative flux emitted by the gas
can be altered by subsequent passage through the scattering medium.
Because of the larger Rayleigh scattering and photon-destruction
cross-sections, the $\lambda$ 1032 line photons undergo much more
interaction. Thus, the simple 2:1 flux ratio which is expected under
nebular conditions can actually be somewhat higher or lower than 2,
depending on phase angle. In particular Schmid finds
f(1032)/f(1038)$<$2 for phase angles where absorption is important and
$>$2 for phase angles where the reflection effect is significant.

\subsection{O~VI Line Flux Variations}

In symbiotic stars, emission line fluxes vary on time scales from as little
as a few weeks to as long as years.  These variations are the result
periodic obscuration due to the binary orbit or of outburst activity of
the hot star.  Many of the same symbiotic systems were observed using
the HUT (Astro-2) and ORFEUS-I and -II instruments in March 1995,
September 1993, and November 1996, respectively.  In order to assure
consistency with Schmid et al. (1999) we will compare the fluxes of
the O~VI lines from the HUT data measured relative to the local linear
continuum in the O~VI region.  For the discussion that follows we are
careful to adopt Orfeus-I and -II fluxes from measurements which are
free of losses due to saturation or misplacement of the target in the
aperature (Schmid et al. 1999).  If we compare the spectra from 1993
to 1996, we see clear variations in the O~VI line fluxes for Z~And,
AG~Dra, V1016~Cyg, and RR~Tel, see Table 9.  In AG~Dra, the 1032\AA\
line flux decreases steadily from 1993 to 1996; the 1038\AA\ flux
decreases by almost 50\% from 1993 to 1995 but does not appear to
change significantly from 1995 to 1996.  In the case of Z~And, the
1032\AA\ line flux remains relatively constant from 1993 to 1995 but
increases by $\sim$40\% from 1995 to 1996 while the 1038\AA\ line flux
increases by $\sim$70\% from 1993 to 1995 and remains unchanged from
1995 to 1996.  For RR~Tel, the fluxes of both O~VI lines decreased
slightly over the period from 1993 to 1996.  In V1016~Cyg, the flux of
the 1032\AA\ line remained constant from 1995 to 1996, while the
1038\AA\ line flux doubled.

Given the differences in both resolution and signal-to-noise between
the HUT and the ORFEUS-I and -II instruments, it is natural to wonder
if these variations in flux are real.  In each study, the observed
O~VI line fluxes are measured relative to the local linear continuum
on either side of the O~VI emission (Schmid et al. 1999).  The
continuum in each object is so weak relative to the O~VI emission that
differences in the choice of the continuum level and the
signal-to-noise in the continuum are not particularly critical for the
determination of the O~VI line fluxes.  The higher resolution ORFEUS
data show no lines in the vicinity of the O~VI emission which could be
blended with the O~VI lines, and hence unresolved, in the lower
resolution HUT data.  Thus, we believe that the observed flux
variations between the HUT and ORFEUS data are real and not an
artifact of either differences in instrumental properties or analysis
technique.

In principle, the observed O~VI line fluxes can vary for a number of
reasons.  Many symbiotic stars show variations in emission line fluxes
with orbital phase (Kenyon 1986) and these variations are associated
with the binary nature of the systems.  Part or all of the line
emitting portion of the ionized nebula surrounding the hot star can be
completely occulted by the red giant, resulting in decreased emission line
fluxes.  Additionally, Shore \& Aufdenberg (1993) have shown that the
outer atmosphere of the red giant star can effectively absorb flux
from many emission lines; such ``atmospheric eclipses" would also
result in flux variations in observed emission lines over the course
of the binary orbital period.  In the models of Schmid (1996) the
simple O~VI flux ratio (2:1 under optically thin conditions) is
sensitive to the orbital phase of the system.  The $\lambda$1032 line
photons undergo much more interaction in the H$^0$ region than the
$\lambda$1038 line photons; this is due to the fact that the shorter
wavelength photons have a 5.2 times larger Rayleigh scattering
cross-section and a 2.3 times greater photon-destruction
cross-section.  Thus, depending on the orbital phase, the flux ratio
f(1032)/f(1038) can sometimes be either less or greater than 2.  This
is the most likely explanation for the variations in the observed flux
ratios for Z~And, V1016~Cyg, and AG~Dra.  Outburst activity also
results in changes in emission line fluxes as well as continuum
emission (Kenyon 1986): many emission lines weaken considerably as the
brightness increases during an outburst and strengthen as the
brightness fades.

The decreasing flux in both O~VI lines of RR~Tel is consistent with
the slow evolution of the RR~Tel spectrum which has been observed over
a number of years (Nussbaumer \& Dumm 1997, Contini \& Formiggini
1999); but, the source of flux variations in the other three
objects is more complicated.  The UV emission lines of O~VI and N~V
should show similar flux variations with orbital phase because they
have similar properties: both are resonance doublets of highly ionized
species and both have wavelengths close to a Lyman transition where
the Rayleigh scattering cross-sections are large (Schmid 1995, 1996;
Schmid et al.  1999).  The N~V 1240\AA\ line in Z~And varies markedly
with orbital phase (Fern{\'a}ndez-Castro et al.  1988, 1995).  In
AG~Dra, the N~V 1240\AA\ line also shows significant variations with
orbital phase (Miko{\l}ajewska et al.  1995, Gonz\'alez-Riestra et
al. 1999).  The HUT observations in March 1995 were obtained after
outburst activity from June to November 1994 and before a secondary
outburst which began in July 1995.  The Orfeus-II observations in 1996
were obtained near the end of an outburst.  The variations in observed
O~VI line flux in AG~Dra from 1993 to 1996 are probably the result of
both outburst activity and orbital variation.  On the other hand,
Z~And was in quiescent state from 1988 to 1997 and V1016~Cyg has been
in quiescence since about 1985, so it would seem that the variations
in O~VI line fluxes are the result of orbital motion in these systems.

\section{Raman Scattered O~VI Lines}

Raman scattering of O~VI photons can occur only in high-excitation
symbiotic systems where O$^{+5}$ ions exist.  To give rise to Raman
scattered emission, a high-excitation source is required to produce
the O~VI photons, and a large H~I column is necessary in order to
provide the scattering source.  Of the total sample of high excitation
systems examined to date, roughly 70\% exhibit the Raman scattered
features at 6825, 7082~\AA\ (Schmid 1992).  Of the six systems in our
study that have O~VI 1032, 1038~\AA\ emission, four (Z~And, V1016~Cyg,
AG~Dra, and RR~Tel) exhibit Raman-scattered O~VI emission, one
(HM~Sge) shows occasional evidence for the scattered lines, and one
(CH~Cyg) has never been seen to exhibit the scattered features.  The
statistics of our (admittedly small) sample are therefore consistent
with Schmid's results (1992).

As mentioned previously, no simultaneous optical observations of
HM~Sge were available for comparison with the HUT far-UV data.  Schmid
et al.  (2000) report that the Raman features have only recently
emerged in the optical spectrum of HM~Sge.  Their optical spectrum
from July 1995 shows weak emission features at $\lambda\lambda$ 6825,
7082.  The data are not fluxed but the measured relative strengths of
the Raman lines EW(6825)=(4.6$\pm$0.3)$\times$EW(7082) are consistent
with the results of Allen (1980).  A lower resolution observation of
HM~Sge taken over a year later with the WIYN is fluxed, see Table 2;
the $\lambda$6825 feature is again weak and the $\lambda$7082 feature
is not visible.  The variation in HM Sge photometry
(ftp://cdsarc.u-strasbg.fr/pub/afoev/) for Astro-2/La Palma/WIYN
observations quoted as a flux ratio is 1.00/0.97/1.03.  We take the
continuum flux of the WIYN observations to be the same as the La Palma
observation, consistent with the photometry.  The observed flux during
the La Palma observation is thus
~1.4$\times$10$^{-14}$~erg~cm$^{-2}$~S$^{-1}$.  Taking E(B-V)=0.65,
0.40 the estimated scattering efficiencies are 0.5 to 0.7\%,
respectively.  As HM~Sge is one of the hottest white dwarf symbiotic
systems, this is consistent with the picture requiring a high density
region of H~I.  In the case of HM~Sge, the H~I is probably nearly
totally ionized.  Schmid et al. (2000) explain the sudden appearance
of the Raman lines by a strong decrease of the absorption of the
far-UV emission lines at Ly$\beta$ and O~VI by dust associated with
the system itself.  This implies that further correction to the Raman
scattering efficiency factor might be necessary.

CH~Cyg also exhibits strong O~VI emission but no Raman features and,
to our knowledge, has never exhibited the Raman lines.  A WIYN
spectropolarimetry observation taken on 9 April 1996 shows a strong
polarization variation in H$\alpha$, but no polarization in excess of
the continuum polarization in the regions of the Raman lines.  In
addition, CH Cyg has been monitored at the Pine Bluff Observatory
({\tt http://www.sal.wisc.edu/HPOL}) for a number of years and neither Raman
line has been observed either in the flux or polarization spectra.
The models of Schmid (1996) and Harries \& Howarth (1997) indicate
that the bulk of the Raman scattered photons are produced in the
region between the hot and cool components of the symbiotic system.
Thus, in order for the Raman scattering process to be effective, the
region between the two stars must contain a high density of H$^0$
scatterers.  CH~Cyg shows P-Cygni features indicating an extensive
outflow of gas and has undergone several outbursts since 1965
(Miko{\l}ajewska, Miko{\l}ajewska, and Khudyakova, 1990; Skopal et
al. 1996).  Why then does CH~Cyg not exhibit the Raman features in the
optical?  Munari et al.  (1996) identified several episodes of dust
condensation in the wind of the red giant and one of these occured during the period spanning 1995 and 1996. According to Munari et al. the
spectroscopic and photometric observations of CH~Cyg suggest that the
dust condensation occurs in a spherically symmetric wind around the
red giant star and that the white dwarf stars lies well outside the
dust.  We suggest then that the absence of the Raman features in
CH~Cyg is, like the case of HM~Sge, due to the absorption of the
far-UV O~VI photons by circumstellar dust in or near the neutral 
scattering region.  Interestingly, CH~Cyg is an S-type symbiotic system whereas HM~Sge is a D-type system so it seems that the supression of the Raman effect by circumstellar dust is not limited to the D-type systems.

\subsection{Extinction Re-examined}

The extinction values for symbiotic stars are notoriously difficult to
derive.  Many of the extinction values obtained from our fits to the
HUT UV data are quite different from the values derived by previous
authors using a variety of methods.  Since the accuracy of the
scattering efficiencies depends sensitively on the extinction
correction, we feel that it is appropriate to re-examine extinction
values at this point.  Table 4 summarizes the extinction estimates of
other researchers and those derived in this study.  The extinction
values adopted for use in line flux corrections for objects which
exhibit the Raman scattering effect are indicated in bold in the
table.  

As noted in $\S$4, data for both CH~Cyg and HM~Sge are consistent with
the presence of dust in or around the symbiotic system.  To derive
extinction estimates appropriate to the O~VI lines, then, it is
necessary to choose a diagnostic arising from material within the same
gas.  Emission lines provide easily measured features and lines from
He~II and [Ne~V] are good candidates.   As the O~VI region is also
likely to be concentrated around the hot star, stellar continuum fits
provide an alternative measurement technique.

As a starting point, we assume Case B conditions and utilize the He II
line ratios to determine extinctions for each object.  Proga et al.
(1996) point out that illumination effects in symbiotics can result in
significant deviations from Case B conditions, making the He~II line
ratios poor reddening diagnostics unless the hot star luminosity and
temperature are well known.  Nonetheless, we believe that the He~II
line ratios can still be used as an effective indicator of reddening
if the results derived from the three line ratios 1085/1640,
1085/4684, 1640/4686 are consistent with their case B values.  We
derive extinctions by comparing the observed He~II line ratios to the
theoretical Case B line ratios of Storey \& Hummer (1995).  We assume
that the emitting gas is optically thin.  For individual objects we
assume nebular conditions as follows: Z~And, T$_e \sim$~20,000~K, N$_e
\sim$~10$^9$~cm$^{-3}$ (Birriel, Espey, Schulte-Ladbeck 1998);
V1016~Cyg T$_e \sim$~20,000~K, N$_e \sim$~10$^7$--10$^8$~cm$^{-3}$
(Nussbaumer \& Vogel 1990); AG~Dra T$_e \sim$~15,000~K, N$_e
\sim$~10$^9$ -- 10$^{10}$~cm$^{-3}$ (Miko{\l}ajewska \& Kenyon 1992);
RR~Tel T$_e \sim$ 20,000~K, N$_e \sim$~10$^6$~cm$^{-3}$ (Espey et
al. 1995a).  In Z~And and V1016~Cyg, the three He~II line ratios yield
consistent reddening values and we conclude that the assumption of
Case B conditions is valid for these objects.  On the other hand, the
derived reddening for RR~Tel and AG~Dra varies considerably depending
on which He~II line ratio is used (Table 4).  The He~II 1085 line
suffers only nominal absorption (less than 5\%) from interstellar
H$_2$, and thus absorption cannot explain the discrepancies among the
derived reddening values.  Our conclusion is that the nebular gas of
both AG~Dra and RR~Tel departs from simple Case B conditions.

Schlegel, Finkbeiner, and Davis 1998 (hereafter SFD) have recently
constructed a full-sky map of the Galactic dust.  Their map is a
reprocessed composite of the {\it COBE}/DIRBE and {\it IRAS}/ISSA maps
and has a resolution of 6'.1 and are shown to predict reddening  with
an accuracy of 16\%.  The new dust map leads to reddening estimates
which are consistent with the Burstein-Heiles maps in most regions of
the sky but are twice as accurate and appear to be much more reliable
in regions of high extinction (SFD).  The authors have made the map
and software for deriving extinctions readily available for general
use.  In Table 4, we have listed the extinctions based on this new
dust map for each of our nine symbiotic stars.

For each object, with the exceptions of CH~Cyg and HM~Sge, we obtain
an extinction value from our fitting of the HUT UV continuum.  A quick
perusal of Table 4 shows that the extinction values derived from a fit
to the HUT UV continuum are generally rather different from the values
obtained by previous fits to the IUE UV continuum (M\"urset et
al. 1991) and other methods.  Our fits to the HUT continuum assume the
extinction curve of Cardelli, Clayton, and Matthis (1989) while
M\"urset et al.  assume the extinction curve of Seaton. These two
extinction parameterizations differ significantly for $\lambda <$
1500\AA.  In addition, M\"urset et al. fit the IUE spectra by eye and
apply no reasonable numerical criterion for the quality of the fit
whereas our results are based on a $\chi^2$-minimization technique.
The differences between our derived extinction values and those of
M\"urset et al.  are most likely a combination of the above mentioned
differences in fitting technique.  However, the results of our
continuum fitting generally agree well the extinction values derived
from the dust maps of SFD.  The results of our continuum fits for
AG~Dra, 0.08$\pm$0.01, and RR~Tel, 0.08$\pm$0.02, respectively, are in
good agreement with previous studies as well as with the recent
results of SFD.

In the case of Z~And, we derive an extinction of 0.21$\pm$0.01 from
our continuum fit and 0.24$\pm$0.03 taking the theoretical Case B He
II line ratios from Storey and Hummer (1995).  While these values are
consistent with one another, they are slightly lower than the values
generally adopted by other groups (Table 4).  Miko{\l}ajewska \&
Kenyon (1996) give a thorough discussion on the various determinations
of the extinction for Z~And.  Since our results are also in agreement
with the extinction estimates from the recent results of SFD, we
assume an extinction of 0.24$\pm$0.03 for Z~And as derived from our
He~II ratios.

Previous estimates for the reddening of V1016~Cyg range from
0.20$\pm$0.10 to 0.30$\pm$0.10, see Table 4.  We derive 0.24$\pm$0.03
from our fit to its HUT UV continuum.  Assuming Case B recombination
conditions (Storey \& Hummer 1995), the He~II line ratios
1640\AA/4686\AA\ and 1085\AA/4686\AA\ yield a reddening of
0.35$\pm$0.04.  Using the dust maps of SFD, the reddening in the
direction of V1016~Cyg is 0.25$\pm$0.04.  We adopt the extinction
derived from our continuum fit for V1016~Cyg, E(B-V)=0.24$\pm$0.03, as
this is more consistent with the SFD results than the higher results
derived using the He~II ratios.

\subsection{Raman Scattering Efficiencies}

The Raman scattering efficiency is defined as the photon ratio of the
Raman scattered line and the initial O~VI line component,
N$_{\mathrm{Raman}}$/N$_{\mathrm{O~VI}}$, i.e. the fraction of emitted
O~VI photons converted to Raman photons.  From the ratios of the
dereddened fluxes, F$_{\lambda 6825}$/F$_{\lambda 1032}$ and
F$_{\lambda 7082}$/F$_{\lambda 1038}$, we derive the photon ratios and
hence the scattering efficiencies for each O~VI line (see Table 10).
The uncertainties in the conversion efficiencies are dominated by the
uncertainties in the extinction values.  RR~Tel and AG~Dra have fairly
well determined extinctions and thus their derived scattering
efficiencies are relatively accurate.  For RR~Tel, V1016~Cyg, and
Z~And, the conversion efficiency for the $\lambda$1032 $\rightarrow$
$\lambda$6825 process is higher than the $\lambda$1038 $\rightarrow$
$\lambda$7082, as a result of the larger scattering cross section for
the former process relative to the latter.  In the case of AG~Dra, the
scattering efficiencies are very similar.  This is in qualitative
agreement with Schmid et al. (1999), whose contemporaneous far-UV and
optical data for AG~Dra in 1995 also yielded a similar scattering
efficiency for both the 1032~\AA\ and 1038~\AA\ line photons.  Given
the difference in orbital phases for the AG~Dra observations in this
study and those of Schmid et al., it is difficult to explain the
similarity of the two Raman processes as a result of orbital effects
(Schmid 1996).  One possible explanation is that the circumstellar
material in the AG Dra system is such that a 1032\AA\ photon is
absorbed or scattered much more than a 1038\AA\ photon, resulting
apparently similar scattering efficiencies for both processes.

Our scattering efficiencies for the Raman process in RR~Tel and Z~And
agree well with those of Schmid et al.\ (1999) while our results for
V1016~Cyg and AG~Dra are quite different.  For V1016~Cyg our
scattering efficiency is $\sim$2 times larger for the $\lambda$1032
$\rightarrow$ $\lambda$6825 process and $\sim$3 times larger for the
$\lambda$1038 $\rightarrow$ $\lambda$7082 process.  In the case of
AG~Dra, our derived efficiencies for both Raman processes are $\sim$3
times smaller than those determined by Schmid et al., though the
similarity in scattering efficiency for both O~VI doublet lines is
consistent with their findings.  For V1016~Cyg and AG~Dra, our derived
scattering efficiencies differ from those of Schmid et al. through our
adoption of a different extinction estimate for each object and our
use of optical spectra contemporaneous with our far-UV data.

The derived scattering efficiencies can be compared to the models of
Schmid (1996).  For the discussion that follows, we assume the
``basic'' model: in this model the photons are initially released as
far-UV O~VI line photons from an extended emission region in the
ionized nebula and may undergo multiple scatterings before escaping.
The scattering efficiency depends strongly on the ionization geometry
as described by the X${_{H^+}}$-parameter defined by Seaquist et
al. (1984).  The dimensionless X${_{H^+}}$-parameter is related to the
ionizing photon luminosity of the hot source L$_H$, the binary
separation p, and the wind parameters of the cool giant ($\dot{M}$,
$v_{\infty}$) according to
\[ X{_{H^+}} \sim pL_H \left (\frac {v_{\infty}}{\dot{M}} \right) ^2 .\]

The scattering efficiencies for RR~Tel and Z~And suggest that about
one in every three emitted O~VI photons interacts with the scattering
region either via Raman scattering, Rayleigh scattering or absorption.
The inferred scattering geometry is a cone-shaped neutral hydrogen
region around the cool giant which is irradiated by an extensive O~VI
emitting region surrounding the hot star, somewhere between the
X$_{\mathrm B3}$ and X$_{\mathrm C3}$ models of Schmid (i.e.
X${_{H^+}}$ between 4.0 and 0.4).  For V1016~Cyg and AG~Dra, which
exhibit similar scattering efficiencies, about 50\% (or roughly one in
two) of the emitted O~VI photons interact with the scattering region.
For V1016~Cyg and AG~Dra the suggested scattering geometry appears to
be quite close to the X$_{\mathrm C3}$ model of Schmid, i.e. a large
neutral hydrogen region irradiated by a much more confined O~VI
emission zone surrounding the hot star (i.e. X${_{H^+}}\sim$0.4).  The
X${_{H^+}}$ parameters derived here are in qualitative agreement with
the results of M\"urset et al.  (1991) in that
0.3$\leq$X${_{H^+}}\leq$4 for these four symbiotic systems.  However,
the Raman scattering efficiency for V1016~Cyg implies that
X${_{H^+}}$ is very close to 0.4 whereas M\"usert et al. derive a much
larger value of X${_{H^+}}\sim$~2.3--3.7.

The models of Schmid (1996) and Harries \& Howarth (1997) show that,
as expected, the production of Raman-scattered O~VI photons is
increased for more extended and dense neutral scattering regions.
Speaking quite generally, our derived Raman efficiencies imply that
V1016~Cyg and AG~Dra have higher mass loss rates than Z~And and
RR~Tel.  Using Figure 12 of Schmid (1996), the derived efficiencies
for AG~Dra and V1016~Cyg yield a mass loss rate of
$\sim$3$\times$10$^{-6}$M$_{\odot}$ and for Z And and RR Tel
$\sim$1$\times$10$^{-6}$M$_{\odot}$.  From radio measurements at
3.6~cm, Seaquist, Krogulec, \& Taylor (1993) report mass loss rates of
1.3$\times$10$^{-5}$M$_{\odot}$ for V1016~Cyg,
6.3$\times$10$^{-6}$M$_{\odot}$ for RR~Tel,
1.8$\times$10$^{-6}$M$_{\odot}$ for Z~And, and
$<2.4\times$10$^{-8}$M$_{\odot}$ for AG~Dra.  These authors point out
that the Wright and Barlow (1975) formula effectively measures the
mass loss associated with the ionized gas only and thus underestimates
the true mass-loss rate.  They find that even for symbiotic stars with
very extended ionized regions, i.e. X$_{H^+}>$1, the Wright-Barlow
formula underestimates mass-loss rates by $\sim$20\%.  For systems
with X$_{H^+}<$1, the mass-loss rate is underestimated by a factor of
two or more.  We suggest that the Raman efficiencies can be used to
provide an independent determination of the X$_{H^+}$ parameter in
symbiotic systems providing another useful method by which mass-loss
rates of cool giants are examined.

\section{O VI Induced Fe~II Bowen Fluorescence}

Early indirect evidence for the far-UV O~VI lines in symbiotic stars
included the presence of Fe~II transitions ($\lambda$1776.56,
$\lambda$1881.20, $\lambda$1884.12) that are selectively photoexcited
by the strongest member of the O~VI resonance doublet at 1031.9\AA\
(Johansson 1988). As is the case with the Raman scattered lines in the
optical, the Fe~II Bowen lines at 1776.56, 1881.20, and 1884.12 \AA\
are found in the spectra of symbiotic systems which exhibit other
high-exitation lines (Johansson 1988 and Fiebelman, Bruhweiler, \&
Johansson 1991, hereafter FBJ).  Weak Fe~II Bowen lines have been
observed in the IUE spectra of Z~And, V1016~Cyg, and RR~Tel
(Feibelman, Bruhweiler, and Johansson 1991).

We searched for the Fe~II Bowen line at 1776~\AA\ in the HUT spectra
of our symbiotic stars.  One of the lines in the region near the
expected location of the Fe~II Bowen line is the Si~II multiplet near
1814~\AA.  The ionization potentials of these two species are quite
similar and these lines should be formed in roughly the same region.
We tied the wavelength of the Fe~II Bowen line at 1776\AA\ to the
theoretical wavelength of the $\lambda$ 1817 member of the Si~II
multiplet and fixed the FWHM of the lines to the resolution of the HUT
as the Bowen lines are narrow (FBJ).

The Fe~II $\lambda$ 1776 line is clearly visible in the HUT data for
Z~And, and is present but extremely weak in V1016~Cyg, and RR~Tel, see
Figure 5.  We did not find evidence for Fe~II 1776~\AA\ flourescent
emission in the HUT data for the six remaining systems (i.e., EG~And,
BF~Cyg, CH~Cyg, AG~Dra, AX~Per, HM~Sge) to a 3$\sigma$ upper limit of
1.0$\times~10^{-13}~\rm~erg~cm^{-2}~s^{-1}$.  The non-detections in
CH~Cyg, AG~Dra, and HM~Sge are consistent with the IUE results of FBJ,
although the authors did report detections of the longer wavelength
fluorescent features in HM~Sge and AG~Dra in the earlier IUE data.

We define the Bowen efficiency, similarly to the Raman efficiency, as
the ratio of the number of Fe~II $\lambda$1776 photons to the
initially emitted number of O~VI~$\lambda$1032 photons.  Table 11
lists the extinction corrected fluxes for the Fe~II~$\lambda$1776.56
line and the efficiency of the Bowen fluorescence mechanism for each
object.  The efficiency of the Bowen mechanism is considerably smaller
than that for the Raman process, less than 2\% in each object.  Our
derived efficiencies for the production of the Fe~II Bowen line at
1776.56\AA\ are consistent with the efficiencies for the Bowen process
in which the C~IV~$\lambda$ 1548.2 line photons photoexcite the Fe~II
1975.5\AA\ line.  Simultaneous IUE fluxes for the C~IV 1548.2\AA\ and
the Fe~II 1975.5\AA\ lines are reported in FBJ for RR~Tel, Z~And, and
V1016~Cyg and the efficiencies derived from these fluxes are also $<$
2\% in each object.

\section{Summary}

From HUT observations and simultaneous or near-simultaneous optical
observations we confirm the contemporaneous presence of the O~VI 1032,
1038\AA\ resonance doublet and the Raman scattered O VI emission at
6825, 7082\AA\ in four of the nine target symbiotic systems, namely
Z~And, V1016~Cyg, AG~Dra, and RR~Tel.  We calculate scattering
efficiencies for each system (Table 9).  Our observations show that in
no case do we observe the Raman scattered lines in the optical without
the simultaneous occurence of the far-UV O~VI emission lines (see
Table) thus giving the strongest support to date that the
optical emission lines at 6825, 7082\AA\ are the product of
Raman-scattered O~VI photons.
  
The existence of strong O~VI and neutral hydrogen in a symbiotic
system is a necessary but insufficient condition for the Raman
scattering process to occur.  Models (Schmid 1996, Harries \& Howarth
1997) indicate that the majority of Raman scattered photons are
produced in the neutral region between the hot star and the red
giant. If the neutral region surrounding the red giant also forms
heavy dust condensation, then the far-UV O~VI photons may be heavily
absorbed and no Raman scattering occurs. This is a strong possibility
for HM Sge which has only recently developed Raman lines in the
optical (Schmid et al. 2000) and for CH~Cyg, which despite evidence for
strong O~VI emission and significant mass loss from the red giant,
does not and apparently never has exhibited the Raman lines.  In
support of this, both objects have apparently experienced several
periods of strong dust condensation around the red giant component
(Schmid et al. 2000, Munari et al. 1996).

The Raman scattering efficiency depends strongly on the geometry of
the neutral hydrogen scattering region and the mass loss rate (Schmid
1996, Harries \& Howarth 1997, Lee \& Lee 1997).  Schmid has derived
efficiencies for various shapes of ionization fronts based on the
X$_{H^+}$-parameterization.  For the four systems in which we observed
the Raman process, the derived Raman scattering efficiencies imply
that X$_{H^+}$~0.4--4.0, in qualitative agreement with M\"urset et al.
(1991) and the previously published ORFEUS results (Schmid et
al. 1999).  The relative mass loss rates of Z~And, V1016~Cyg, AG~Dra,
and RR~Tel based on their Raman efficiencies are not in general
agreement with those derived from radio measurements (Seaquist et
al. 1993).  As both the Raman line intensity and polarization profiles
are very sensitive to the mass loss rate (Harries \& Howarth 1997),
detailed studies of these profiles may provide another method to
examine mass loss in symbiotic systems.

The presence of Fe~II transitions ($\lambda$1776.56, $\lambda$1881.20,
$\lambda$1884.12) that are selectively photoexcited by the strongest
member of the O~VI resonance doublet at 1031.9\AA\ was presented as
early indirect evidence for the presence of the the far-UV O~VI lines
at 1032, 1038~\AA\ (Johansson 1988).  In our small sample of symbiotic
systems, we find no case in which the Fe II 1776.56\AA\ line is
present without the simultaneous occurence of the O~VI line at
1032\AA.  Thus, the HUT data provide the first observational support
for the Bowen fluoresence origin of the Fe II 1776.56\AA\ line.
Furthermore, the data enable us to make the first estimates of the
efficiency for the Bowen fluorescence mechanism responsible for the
production of the 1776.56\AA\ in the symbiotic systems Z~And, RR~Tel,
and V1016~Cyg.  The derived efficiency is less than 2\% in each
system.

The prerequisites for the Raman scattering process and the Bowen
fluoresence process are quite similar.   Both require the presence of
strong O~VI emission and sufficient amounts of H~I and Fe~II.
Additionally, H~I and Fe~II have similar ionization potentials,
13.6~eV and 16.2~eV, respectively.  Therefore the Raman scattering and
Bowen fluorescence processes must both occur in roughly the same
region,  close to the red giant.    Observations from our small sample
of symbiotic systems support this: each object which exhibits the
Raman scattering process (i.e. Z~And, V1016~Cyg, AG~Dra, HM~Sge,
RR~Tel) also exhibits some or all of the Fe~II Bowen lines (FBJ).
CH~Cyg shows strong O~VI emission but does not exhibit either the
Raman lines or the Bowen lines.  It might make an interesting study to
see if the mutual presence of the Fe~II Bowen and Raman lines holds
over a larger sample of symbiotics and also if the efficiences of the
two processes are correlated in any way.

The authors wish to thank Drs. Kenyon and Miko{\l}ajewska and
Drs. Schild \& Schmid for providing the optical data for this study.
The WIYN data were obtained in a collaboration with Drs. J. Johnson
and C. Anderson.  We also wish to thank Dr.~Dixon for supplying data
in advance of publication and Dr. McCandliss for generating the H$_2$\
optical depth templates from the laboratory molecular hydrogen data.
We thank the observers of the AFOEV for making their data public and
hence aiding in the calibration of our spectroscopic data.  This work
was supported by Astro-2 Guest Investigator grant NAG8-1049 to the
University of Pittsburgh.  Jennifer Birriel acknowledges the support
of NASA Graduate Student Researcher Grant NGT5-50174.

\newpage

\begin{figure}[ht]
\centerline{\psfig{file=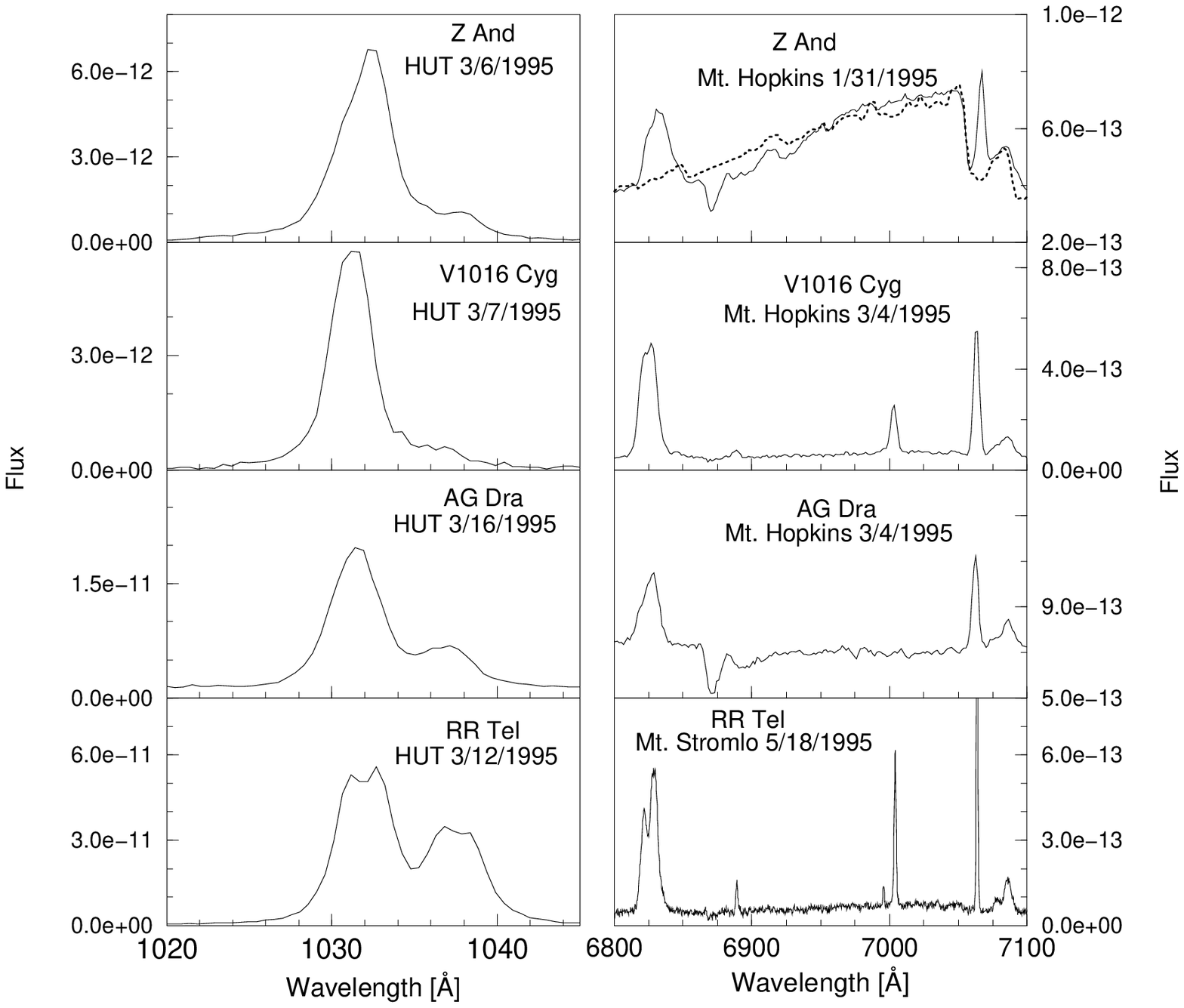}}
\caption{HUT far-UV and ground-based optical data for objects exhibiting both the
direct and Raman scattered O~VI lines. Fluxes are in units of 
ergs~cm$^{-2}$~s$^{-1}$~\AA$^{-1}$. Dotted lines represent our fits to the
continuum of the M giant star.}
\end{figure}

\clearpage

\begin{figure}[ht]
\centerline{\psfig{file=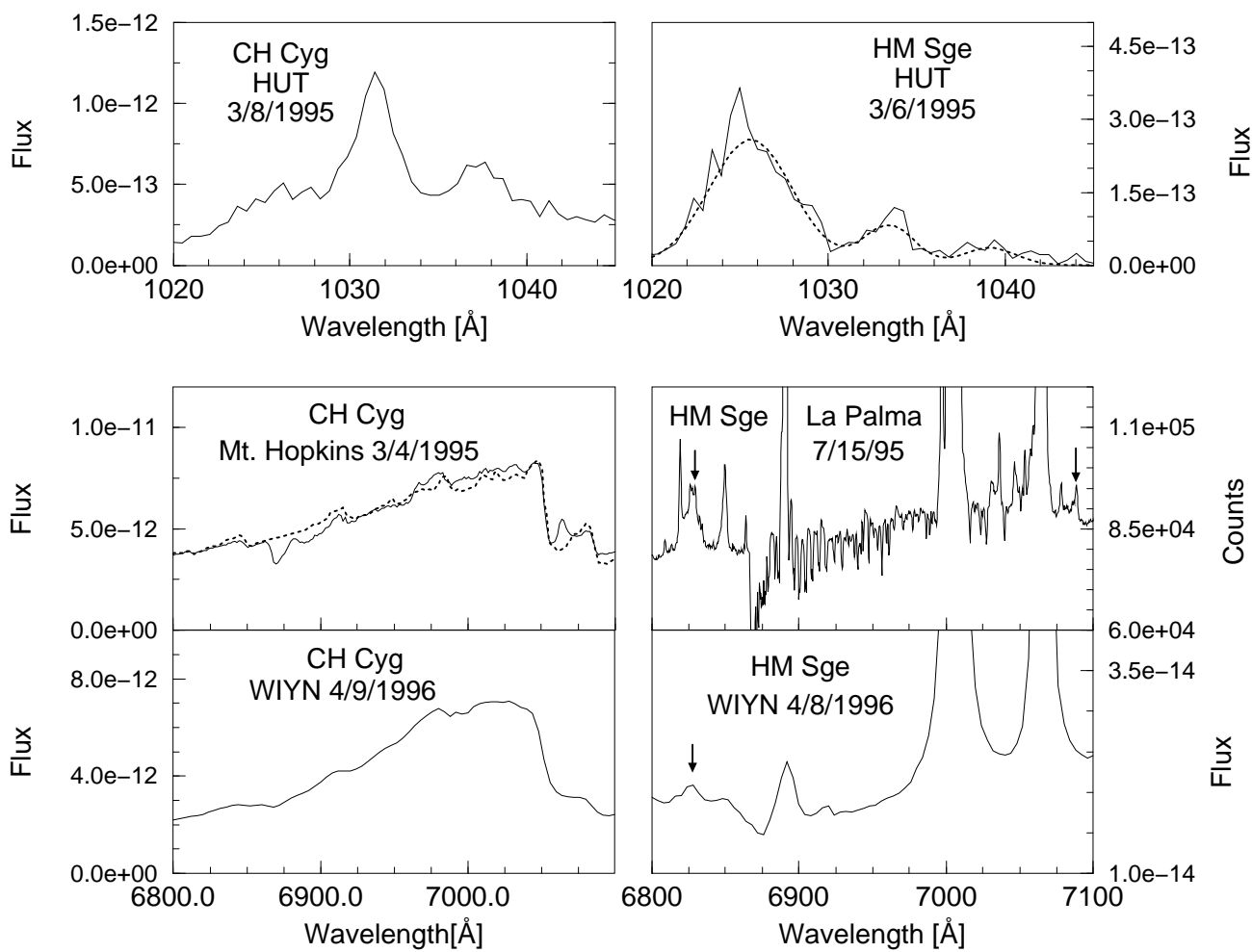}}
\caption{HUT and optical data for objects showing direct O~VI lines in the
far-UV and no Raman lines or extremely weak Raman lines (indicated by arrows) 
in the optical. Fluxes are in units of ergs~cm$^{-2}$~s$^{-1}$~\AA$^{-1}$. The 
emission feature near $\lambda$ 1026\AA\ in the spectrum of HM~Sge is 
terrestrial Ly $\beta$ airglow and the dotted lines represent our best fit to 
the Ly $\beta$ and the weak O~VI lines.}
\end{figure}

\clearpage

\begin{figure}[ht]
\centerline{\psfig{file=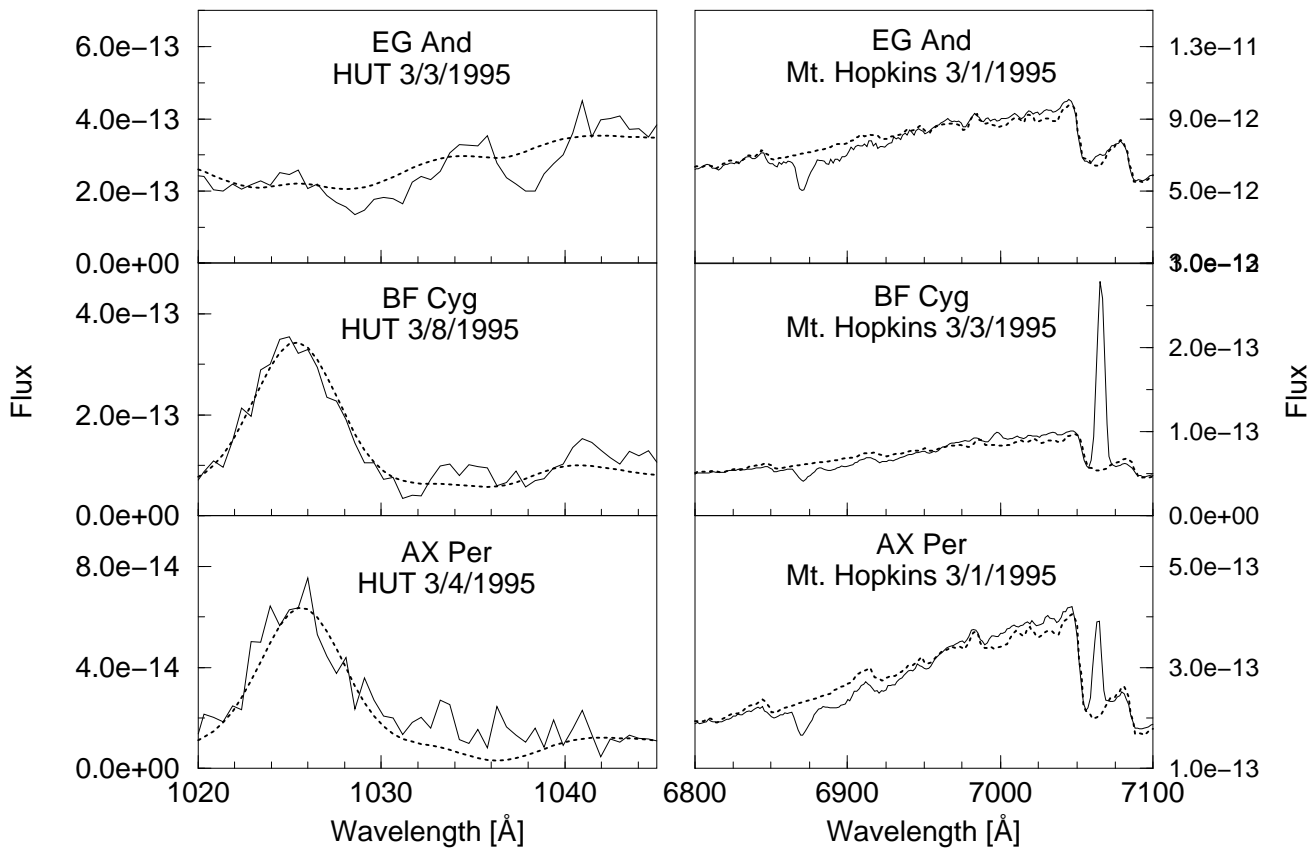}}
\caption{HUT and optical data for objects showing neither the far-UV O~VI lines
nor the optical emission.  Fluxes are in units of 
ergs~cm$^{-2}$~s$^{-1}$~\AA$^{-1}$.  The emission
feature near $\lambda$ 1026\AA\ in the spectra of BF~Cyg and AX~Per is 
terrestrial Ly $\beta$ airglow. Dotted lines represent our fits to the 
continuum.}
\end{figure}

\clearpage
\begin{figure}[ht]
\centerline{\psfig{file=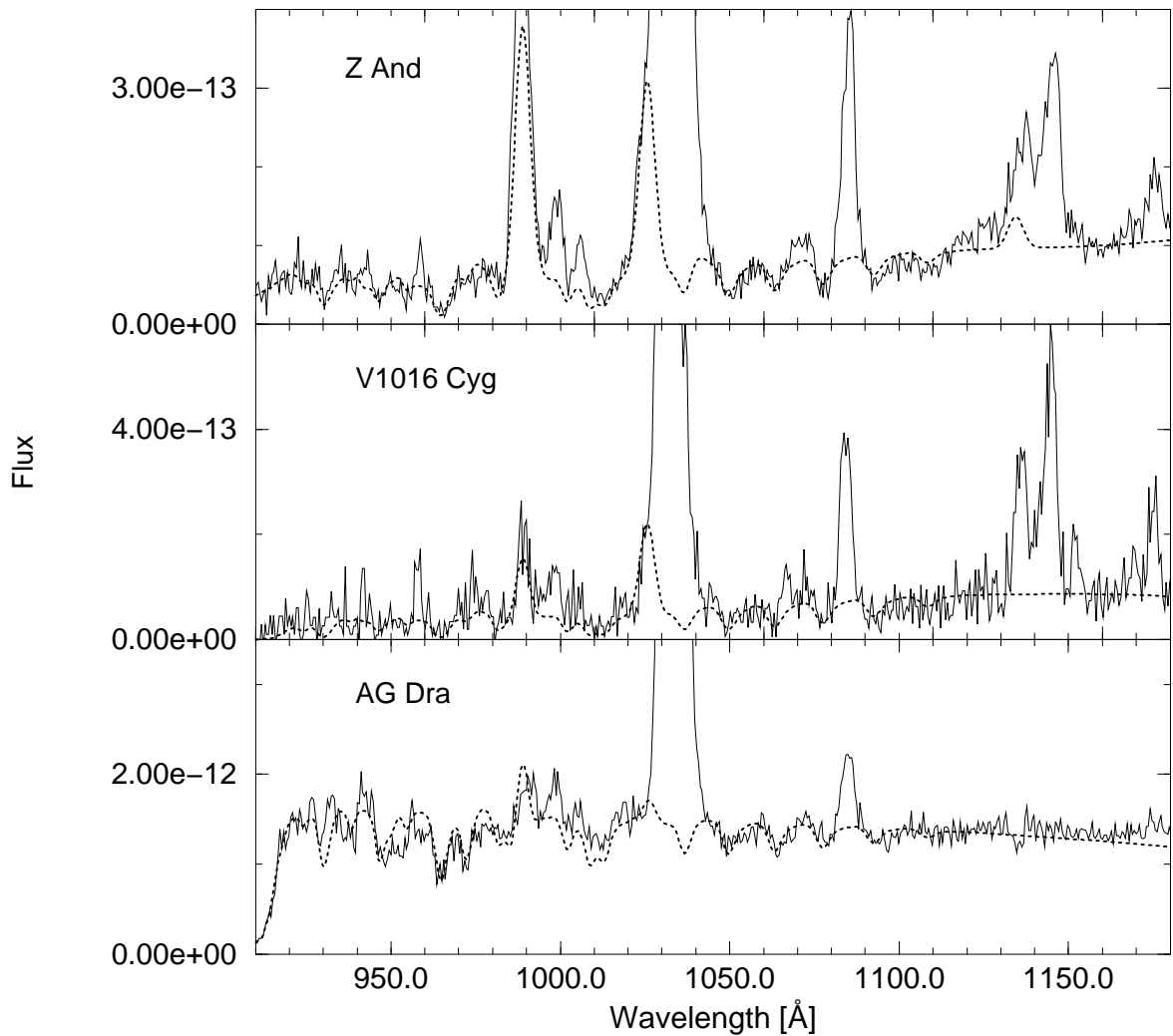}}
\caption{Far-UV continuum fits for Z~And, V1016~Cyg, and AG~Dra.  The fits 
(dotted lines) consist of a blackbody spectrum reddened with the extinction 
curve of CCM (1989) and absorbed by interstellar H~I and H$_2$.  In addition,
we included terrestrial airglow from N~I (weakly visible at $\lambda$ 913\AA),
O~I (989\AA), 
and H~I ($\lambda$ 939 \AA,  $\lambda$ 950\AA, 973\AA, 
$\lambda$ 1026\AA).  Fluxes are in units of 
ergs~cm$^{-2}$~s$^{-1}$~\AA$^{-1}$. }
\end{figure}

\clearpage
\begin{figure}[ht]
\centerline{\psfig{file=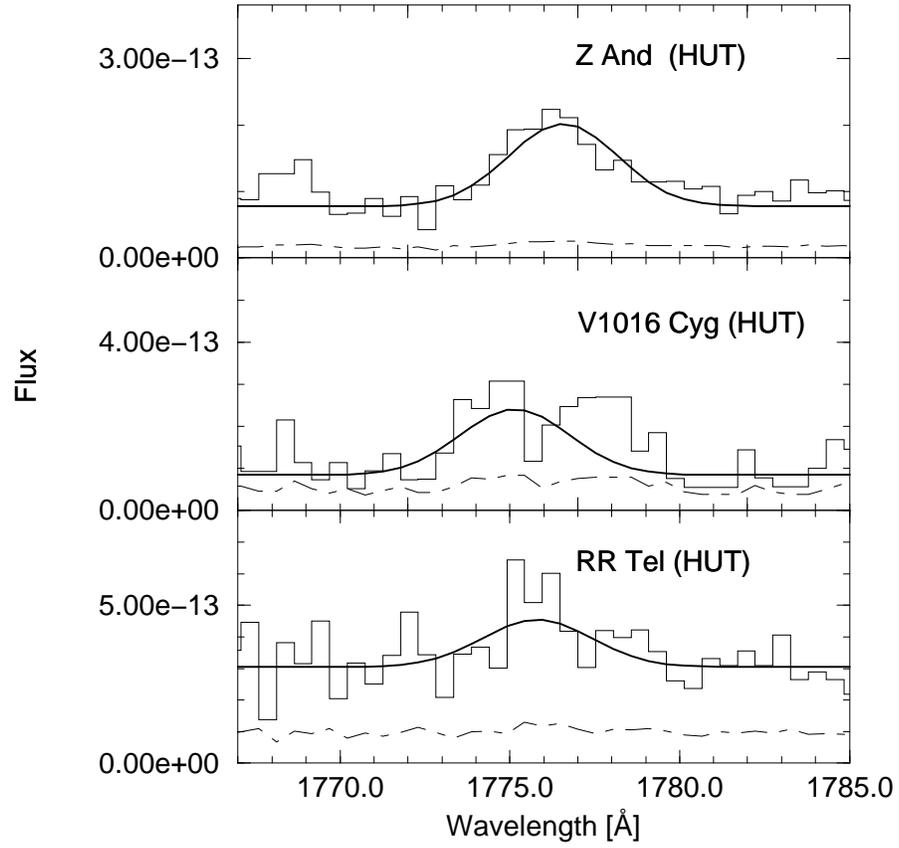}}
\caption{HUT data showing the weak Fe~II Bowen Fluorescence lines at 
$\lambda$ 1776. Fluxes are in units of ergs~cm$^{-2}$~s$^{-1}$~\AA$^{-1}$.
The thick solid lines show our best fit to the data. The thick dot-dash line 
is the error array for the data.}
\end{figure}

\newpage 

\begin{table}
\begin{center}
\centerline{Table 1a} \centerline{HUT Far UV Spectra}
\begin{tabular}{lcccccc} \tableline\tableline
\multicolumn{1}{l}{Target}& \multicolumn{1}{c}{Start Time \tablenotemark{a}}& 
\multicolumn{1}{c}{Exposure}& \multicolumn{1}{c}{Door}& \multicolumn{1}{c}{Slit Diameter}&
\multicolumn{1}{c}{Photometric}& \multicolumn{1}{c}{S/N Near} \\
\multicolumn{1}{l}{}& \multicolumn{1}{c}{(UT)}& \multicolumn{1}{c}{(s)}& 
\multicolumn{1}{c}{(cm$^2$)}& \multicolumn{1}{c}{(arcsec)}& \multicolumn{1}{c}{Correction}&
\multicolumn{1}{c}{O~VI Lines} \\ \tableline

EG~And&      Mar 15 06:29:55&   1328&    5120&      20&    1.0343&  10\\
 & & & & \\
Z~And&      Mar 6 01:21:44&     1370&    5120&      20&    1.0091&  10\\
 & & & & \\
BF~Cyg&     Mar 8 03:37:02&      796&     5120&      20&    1.2287&  8\\
 & & & & \\
CH~Cyg&      Mar 13 02:39:55&   1082&     5120&      20&    1.0993&  10\\
 & & & & \\
V1016~Cyg&    Mar 6 23:58:58&    664&      2560&      12&    1.6483&  4\\
 & & & & \\
AG~Dra&      Mar 16 19:14:18&    764&      2560&      20&    1.0057&  15\\
 & & & & \\
AX~Per&      Mar 4 02:15:59&     548&       5120&      20&    1.5631&  4\\
 & & & & \\
HM~Sge&     Mar 6 10:24:34&     1026&      5120&      20&    1.0817&  1\\
 & & & & \\
RR~Tel&      Mar 12 14:21:53&    994&       750&      20&    1.0167&  10\\
& & & & \\ \tableline
\end{tabular}
\tablenotetext{a}{All observations obtained during orbital day.} 
\end{center}
\end{table} 

\newpage

\begin{table}
\begin{center}
\centerline{Table 1b}
\centerline{Optical Spectra}
\begin{tabular}{lcccc} \tableline\tableline
\multicolumn{1}{c}{Target}& \multicolumn{1}{c}{Date}& \multicolumn{1}{c}{Telescope \tablenotemark{a}}&
\multicolumn{1}{c}{Photometric}& \multicolumn{1}{c}{Reference} \\ 
\multicolumn{1}{l}{}& \multicolumn{1}{c}{}& \multicolumn{1}{c}{}&
\multicolumn{1}{c}{Scale Factor\tablenotemark{b}}& \multicolumn{1}{c}{} \\ \tableline

EG~And&      Jan 3 1995&   Mt. Hopkins&   1.70&   Miko{\l}ajewska \\
& & & & \& Kenyon (1996) (MK)\\
Z~And&       Jan 31 1995&  Mt. Hopkins&   &   MK  \\

BF~Cyg&      Mar 3 1995&   Mt. Hopkins&   &   MK  \\

CH~Cyg&      Apr 3 1995&   Mt. Hopkins&   &   MK  \\
      &      Apr 9 1996&   WIYN&          &       \\

V1016~Cyg&   Apr 3 1995&   Mt. Hopkins&   &   MK  \\

AG~Dra&      Apr 3 1995&   Mt. Hopkins&   5.10&   MK  \\

AX~Per&      Jan 3 1995&   Mt. Hopkins&   2.36&   MK  \\

HM~Sge&      Jul 15 1995&  La Palma&       &   Schmid \& Schild 1997b \\
      &      Apr 8 1996&   WIYN&           &      \\

RR~Tel&      May 18 1995&  Mt. Stromlo&    &  Espey et al. 1995a  \\ \tableline

\end{tabular}
\tablenotetext{a}{Mt. Hopkins Observatory 1.5m telescope; La Palma Observatory 4.2m telescope; 
WIYN: Wisconsin Indiana Yale NOAO 3.5m telescope;
Mt. Stromlo Observatory 1.88m telescope} 
\tablenotetext{b}{For observations affected by clouds, see \S2.} 
\end{center} 
\end{table}

\newpage

\begin{table}
\begin{center}
\centerline{Table 2}
\centerline{Observed Fluxes (erg cm$^{-2}$ s$^{-1}$) of Direct \& }
\centerline{Raman Scattered O VI Lines in Selected Symbiotic Stars}
\begin{tabular}{lcccc} \tableline\tableline

\multicolumn{1}{l}{}& \multicolumn{2}{c}{O VI \tablenotemark{a}}&
\multicolumn{2}{c}{Raman Scattered}\\ \tableline
\multicolumn{1}{l}{Object}& \multicolumn{1}{c}{1032}& \multicolumn{1}{c}{1038}&
\multicolumn{1}{c}{6825}& \multicolumn{1}{c}{7082} \\ \tableline

EG~And&  $<$ 3.0E-14&  $<$ 2.0E-14&  $<$ 6.0E-14& $<$ 1.5E-13 \\
Z~And&  2.6E-11& 4.2E-12& 4.2E-12& 1.3E-12 \\
BF~Cyg& $<$ 1.5E-14&  $<$ 1.5E-14&  $<$ 1.1E-15&  $<$ 2.5E-15 \\
CH~Cyg& 3.1E-12& 1.6E-12& $<$ 4.0E-14& $<$ 1.0E-14 \\
V1016~Cyg& 1.9E-11& 2.1E-12& 6.6E-12& 1.1E-12\\
AG~Dra& 7.1E-11& 2.2E-11& 4.5E-12& 1.9E-12\\
AX~Per&  $<$ 5.1E-14& $<$ 5.4E-14& $<$ 1.3E-15& $<$ 1.9E-14 \\
HM~Sge \tablenotemark{b}& 3.1E-13& 1.4E-13& 3.7E-14& $<$5.1E-14\\
RR~Tel& 2.3E-10& 1.4E-10& 5.4E-12& 1.0E-12\\ \tableline

\end{tabular}
\tablenotetext{a}{O~VI lines measured relative to a local linear continuum.}
\tablenotetext{b}{WIYN observation on 8 April 1996.} 
\tablecomments{Where lines are absent, we give the 3$\sigma$ upper limit 
on fluxes. UV line fluxes are accurate to $\sim$6\%. Optical line fluxes good
to $\sim$10\%.}
\end{center}
\end{table}

\newpage
\begin{table}
\begin{center}
\centerline{Table 3}
\centerline{Hot Star Zanstra Temperatures.}
\begin{tabular}{lll} \tableline\tableline
\multicolumn{1}{l}{}& \multicolumn{2}{c}{T$_{WD}$ (K) \tablenotemark{a}} \\
\multicolumn{1}{l}{Object}& \multicolumn{1}{l}{This Work}& 
\multicolumn{1}{l}{M\"urset et al. 1991}\\ \tableline
EG~And&  76,000&  80,000 (Nov. 1990) \\
Z~And&   110,000&  130,000 (Nov. 1987) \\
BF~Cyg&  64,000& 55,000 (July 1986)\\
      
CH~Cyg&  58,000\tablenotemark{b}&  80,000 (Nov. 1988) \\
    
V1016~Cyg&  135,000&  145,000K (Nov. 1987) \\
         
AG~Dra&  86,000 \tablenotemark{b}&  125,000 (June 1985)\\

AX~Per&  89,000&  105,000 (Oct. 1984)\\
      
HM~Sge&  165,000&  200,000 (Oct. 1990)\\
      
RR~Tel&  130,000&  140,000 (June 1983)\\  \tableline
\end{tabular}
\tablenotetext{a}{Uncertainties of $\pm$20\% } 
\tablenotetext{b}{Outburst, see text.} 
\end{center} 
\end{table}

\newpage
\begin{table}
\begin{center}
\centerline{Table 4}
\centerline{Extinction values from the literature.}
\begin{tabular}{llll} \tableline\tableline
\multicolumn{1}{l}{Object}& \multicolumn{1}{l}{E(B-V)}& \multicolumn{1}{l}{Method}& 
\multicolumn{1}{l}{Reference} \\ \tableline
EG~And&  0.10$\pm$0.01& UV continuum fit& this work \\
      &  0.09$\pm$0.01& Dust maps& SFD \tablenotemark{a} (1998) \\
      &  0.05&  Shape of UV continuum&  M\"urset et al. (1991) \\
Z~And&   {\bf 0.24$\pm$0.03}&  He II 1640/4686, 1085/4686& Birriel et al. (1998)\\
     &   0.21$\pm$0.01& UV continuum fit& this work\\
     &   0.21$\pm$0.03& Dust maps& SFD (1998)\\
     &   0.30&  2200\AA\ feature&  M\"urset et al. (1991) \\
     &   0.2-0.3& H~I Balmer lines& Miko{\l}ajewska \& Kenyon (1996) \\
BF~Cyg&  0.24$\pm$0.01& UV continuum fit& this work \\
      &  0.29$\pm$0.05& Dust maps& SFD (1998) \\
      &  0.35&  2200\AA\ feature&  M\"urset et al. (1991)\\
      &  0.40$\pm$0.05& UV continuum fit& Miko{\l}ajewska et al. (1989) \\
CH~Cyg&  0.08$\pm$0.01& Dust maps& SFD 1998 \\
      &  0.10&  Shape of UV continuum&  M\"urset et al. (1991) \\
      &  0.00&  absence of 2200\AA\ feature& Miko{\l}ajewska et al. (1988)\\
V1016~Cyg& {\bf 0.24$\pm$0.03}& UV continuum fit& this work\\
         & 0.35$\pm$0.04& He II 1640/4686, 1085/4686& this work \\
         & 0.25$\pm$0.04& Dust maps& SFD (1998)\\
         & 0.40& 2200\AA\ feature&  M\"urset et al. (1991) \\
         & 0.17$\pm$0.02& UV continuum fit& Kenyon \& Webbink (1984)\\
         & 0.20$\pm$0.10& [Ne V] 1575/2973 ratio& Nussbaumer \& Schild (1981)\\
         & 0.28& 2200\AA\ feature& Nussbaumer \& Schild (1981) \\  \tableline
\end{tabular}
\end{center} 
\end{table}

\newpage
\begin{table}
\begin{center}
\centerline{Table 4, cont.'d}
\centerline{Extinction values from the literature.}
\begin{tabular}{llll} \tableline\tableline
\multicolumn{1}{l}{Object}& \multicolumn{1}{l}{E(B-V)}& 
\multicolumn{1}{l}{Method}& 
\multicolumn{1}{l}{Reference} \\ \tableline
AG~Dra&  {\bf 0.08$\pm$0.01}& UV continuum fit& this work\\
      &  0.07$\pm$0.01& He II 1640/4686& this work \\
      &  0.12$\pm$0.02& He II 1085/4686& this work \\
      &  0.15$\pm$0.02& He II 1085/1640& this work \\
      &  0.11$\pm$0.01& Dust maps& SFD (1998)\\
      &  0.00&  Shape of UV continuum&  M\"urset et al. (1991) \\
      &  0.06$\pm$0.01&  He~II 1640/4686&   Miko{\l}ajewska et al. (1995) \\
      &  0.05$\pm$0.01&  UV continuum fit&  Miko{\l}ajewska et al. (1995) \\
AX~Per&  0.25$\pm$0.03& UV continuum fit& this work \\
      &  0.22$\pm$0.03& Dust maps& SFD (1998) \\
      &  0.32$\pm$0.02&  He~II 1640/4686&   Miko{\l}ajewska  \& Kenyon (1992)\\
      &  0.25&  2200\AA\ feature&  M\"urset et al. (1991)\\
      &  0.28&  H~I Balmer lines&  Blair et al. (1983) \\
HM~Sge&  0.65$\pm$0.10& Dust maps& SFD (1998) \\
      &  0.65&  2200\AA\ feature&   M\"urset et al. (1991)\\
      &  0.56$\pm$0.14&  H~I Balmer lines&  Pacheco et al. (1989) \\
      &  0.40&  H~I Balmer lines&  Blair et al. (1981)\\
RR~Tel&  {\bf 0.08$\pm$0.02}&  Ne V 1575/2973&  Espey et al. (1996), this work\\
      &  0.05$\pm$0.01& Dust maps& SFD (1998) \\
      &  0.10&  Shape of UV continuum&  M\"urset et al. (1991)\\
      &  0.10&  He II recombination lines& Penston et al. (1983)\\ \tableline
\end{tabular}
\tablecomments{Adopted extinctions appear in bold.}  
\tablenotetext{a}{Schlegel, Finkbeiner, \& Davis.}
\end{center} 
\end{table}

\newpage

\begin{table}
\begin{center}
\centerline{Table 5}
\centerline{Optical Spectral Types for Cool Components}
\begin{tabular}{lcc} 
\multicolumn{1}{l}{Object}&  \multicolumn{1}{c}{This Study/Orbital Phase}&
\multicolumn{1}{c}{Kenyon \& Fernandez-Castro 1987}\\ \tableline \tableline
EG~And&  M2.9$\pm$0.5 III/0.99&  M2.4$\pm$0.3 III\\
Z~And&  M4.1$\pm$0.5 III/0.33&  M3.5$\pm$1 III\\
BF~Cyg&  M4.2$\pm$0.5 III/0.88&  M5$\pm$1 III\\
CH~Cyg&  M5.0$\pm$0.5 III/0.62&  M6.5$\pm$0.3 III\\
AX~Per&  M4.7$\pm$0.5 III/0.17&  M5.2$\pm$0.4 III\\
\tableline
\end{tabular}
\end{center}
\end{table}

\newpage

\begin{table}
\begin{center}
\centerline{Table 6}
\centerline{Model Continuum Parameters}
\begin{tabular}{lccccc} 
\multicolumn{1}{l}{Object}&  \multicolumn{1}{c}{T$_{\rm WD}$ \tablenotemark{a}}&
\multicolumn{1}{c}{T$_{Neb}$ \tablenotemark{b}}&   \multicolumn{1}{c}{E(B-V)}&
\multicolumn{1}{c}{log N(HI)}&  \multicolumn{1}{c}{log N(H$_2$)} \\ \tableline \tableline

Z~And&  110,000K& 20,000 K& 0.21$\pm$0.01& 21.30$\pm$0.08& 
20.06$\pm$0.05  \\
V1016Cyg& 135,000& 10,000 K& 0.24$\pm$0.03& 21.79$\pm$0.82& 20.27$\pm$0.07 \\
AG~Dra&  125,000& 15,000 K&  0.08$\pm$0.01& 20.80$\pm$0.26& 
19.70$\pm$0.04 \\
\tableline
\end{tabular}
\tablenotetext{a}{Blackbody temperature of white dwarf, fixed, see
 text.}
\tablenotetext{b}{Nebular electron temperature; fixed, 
see text.}
\end{center}
\end{table}

\newpage

\begin{table}
\begin{center}
\centerline{Table 7}
\centerline{H$_2$ Absorption Correction Factors}
\begin{tabular}{lccccc} \tableline \tableline
\multicolumn{1}{l}{}& \multicolumn{1}{c}{}& \multicolumn{1}{c}{O~VI Line Width \tablenotemark{a}}&
\multicolumn{1}{c}{V$_{Rad}$ \tablenotemark {b}}& 
\multicolumn{2}{c}{Transmission Coefficient} \\ 
\multicolumn{1}{l}{Object}&  \multicolumn{1}{c}{log N(H$_2$)}&
\multicolumn{1}{c}{(km s$^{-1}$)}&  \multicolumn{1}{c}{(km s$^{-1}$)}&
\multicolumn{1}{c}{1031.928}& \multicolumn{1}{c}{1037.617}  \\ \tableline 
Z~And&     20.06$\pm$0.05&   70&  12&   0.976$\pm$0.003&  0.381$\pm$0.042 \\
V1016Cyg&  20.27$\pm$0.07&   70& $-$52&   0.961$\pm$0.006& 0.248$\pm$0.049 \\
AG~Dra&    19.70$\pm$0.04&   70& $-$138&   0.977$\pm$0.002& 0.618$\pm$0.023 \\
\tableline
\end{tabular}
\tablecomments{The relative velocity of the absorbing H$_2$ cloud and the symbiotic system is
assumed to be 0.5$|V_{Rad}|$, see text.} 
\tablenotetext{a}{Schmid et al. 1998} 
\tablenotetext{b}{Heliocentric radial velocity of object $\pm$4~km~s$^{-1}$(Ivison, Bode, Meaburn 1994).}
\end{center}
\end{table}

\newpage

\begin{table}
\begin{center}
\centerline{Table 8}
\centerline{O VI line fluxes (in 10$^{-12}$ erg cm$^{-2}$ s$^{-1}$) relative 
to H$_2$ Absorbed Continuum}
\begin{tabular}{lccccccc} \tableline \tableline
\multicolumn{1}{l}{}& \multicolumn{2}{c}{Observed}&
\multicolumn{1}{c}{}& \multicolumn{1}{c}{E(B-V)}& 
\multicolumn{2}{c}{Corrected \tablenotemark{a}}& \multicolumn{1}{c}{} \\ 
\multicolumn{1}{l}{Object}&  \multicolumn{1}{c}{1032}&
\multicolumn{1}{c}{1038}& \multicolumn{1}{c}{$\frac{F(1032)}{F(1038)}$}& 
\multicolumn{1}{c}{(mag)}& \multicolumn{1}{c}{1032}& 
\multicolumn{1}{c}{1038}& \multicolumn{1}{c}{$\frac{F(1032)}{F(1038)}$}  \\ \tableline 
Z~And&     26$\pm$1&  4.1$\pm$0.2& 6.3& 0.24$\pm$0.03&  
730$\pm$300&   280$\pm$120& 2.6 \\
V1016Cyg&  19$\pm$1&  2.1$\pm$0.2& 9.0& 0.24$\pm$0.03&
550$\pm$220&  220$\pm$100& 2.5 \\
AG~Dra&  70$\pm$3& 23$\pm$1& 3.0& 0.08$\pm$0.02& 
220$\pm$30&  100$\pm$15& 2.2   \\
\tableline
\end{tabular}
\tablenotetext{a}{Absorption and extinction corrected. Quoted error 
includes uncertainty in absorption and extinction.} 
\end{center}
\end{table}

\newpage
\begin{table}
\begin{center}
\centerline{Table 9}
\centerline{O~VI Line Flux Variations}
\begin{tabular}{lccc} 
\multicolumn{4}{l}{Observed Flux in (10$^{-12}$ erg cm$^{-2}$ s$^{-1}$) \tablenotemark{a} 
/ Orbital Phase}\\ 
\tableline\tableline
\multicolumn{1}{l}{}& \multicolumn{1}{c}{1993 \tablenotemark{b}}& \multicolumn{1}{c}{1995}&
\multicolumn{1}{c}{1996 \tablenotemark{b}}\\ \tableline
\multicolumn{4}{c}{Z~And} \\
1032\AA & 25$\pm$1 / 0.67& 26$\pm$1 / 0.39& 35$^{+7}_{-4}$ / 0.21 \\
1038\AA & 2.5$\pm$0.3 / 0.67& 4.2$\pm$0.2 / 0.39& 4.2$^{+0.8}_{-0.4}$ / 0.21 \\
\multicolumn{4}{c}{V1016~Cyg} \\
1032\AA &  NA \tablenotemark{c}& 19$\pm$1& 19$\pm$1 \\
1038\AA &  NA \tablenotemark{c}& 2.1$\pm$0.1& 1.0$\pm$0.1 \\
\multicolumn{4}{c}{AG~Dra} \\
1032\AA & 95$\pm$5 / 0.18& 72$\pm$4 / 0.66& 62$^{+12}_{-6}$ / 0.28 \\
1038\AA & 42$\pm$4 / 0.18& 23$\pm$1 / 0.66& 21$^{+4}_{-2}$ / 0.28 \\
\multicolumn{4}{c}{RR~Tel}\\
1032\AA & 256$\pm$5& 232$\pm$11& 218$\pm$22 \tablenotemark{d} \\
1038\AA & 154$\pm$5& 143$\pm$7&  132$\pm$13 \tablenotemark{d} \\  \tableline
\end{tabular}
\tablenotetext{a}{Measured relative to a local linear continuum.}
\tablenotetext{b}{Orfeus-I and -II, Schmid et al. 1999} 
\tablenotetext{c}{No spectrum available. } 
\tablenotetext{d}{21 Nov 1996 Echelle spectrum observation.} 
\end{center}
\end{table}

\newpage

\begin{table}
\begin{center}
\centerline{Table 10}
\centerline{Raman Scattering Efficiencies}
\begin{tabular}{lcccccc} 
\multicolumn{7}{l}{Extinction corrected O VI and Raman line fluxes in
(10$^{-12}$erg cm$^{-2}$ s$^{-1}$)} \\ 
\multicolumn{7}{l}{Raman efficiency is the ratio of observed 
Raman photons to {\it initial} } \\  
\multicolumn{7}{l}{number of O~VI photons.} \\  \tableline \tableline
\multicolumn{1}{l}{}&  \multicolumn{2}{c}{O VI \tablenotemark{a}}&
\multicolumn{2}{c}{Raman}& 
\multicolumn{2}{c}{N$_{\mathrm{Raman}}$/N$_{\mathrm {O~VI}}$ (\%)} \\ 
\multicolumn{1}{l}{Object}& \multicolumn{1}{c}{1032}& \multicolumn{1}{c}{1038}&
\multicolumn{1}{c}{6825}& \multicolumn{1}{c}{7082}&
\multicolumn{1}{c}{N$_{6825}$/N$_{1032}$}&
\multicolumn{1}{c}{N$_{7082}$/N$_{1038}$} \\ \tableline 
Z~And&  730$\pm$300& 280$\pm$120& 7.3$\pm$0.9& 2.2$\pm$0.5& 7$\pm$3& 5$\pm$2 \\
V1016~Cyg& 550$\pm$220& 220$\pm$100& 14.4$\pm$2.0& 2.2$\pm$0.3& 15$\pm$6& 6$\pm$3 \\
AG~Dra& 220$\pm$30& 100$\pm$15& 5.4$\pm$0.6& 2.2$\pm$0.2&  
14$\pm$2& 13$\pm$2 \\
RR~Tel& 700$\pm$190& 430$\pm$120& 6.7$\pm$1.0& 1.3$\pm$0.3&  
6.0$\pm$1.9& 2.0$\pm$0.7 \\ \tableline
\end{tabular}
\tablenotetext{a}{Includes corrections for H$_2$ absorptions, see Table 7.}
\end{center}
\end{table}

\newpage

\begin{table}
\begin{center}
\centerline{Table 11}
\centerline{Bowen Fluorescence Efficiencies}
\begin{tabular}{lccc} 
\multicolumn{4}{l}{Extinction corrected O VI and Fe II Bowen line fluxes in units of} \\
\multicolumn{4}{l}{10$^{-13}$erg cm$^{-2}$ s$^{-1}$. The photon ratio 
N$_{\mathrm{Fe II}}$/N$_{\mathrm {O~VI}}$ } \\
\multicolumn{4}{l}{is the Bowen fluoresence efficiency.} \\

\multicolumn{1}{l}{Object}& \multicolumn{1}{c}{F(O VI $\lambda$ 1032)}& 
\multicolumn{1}{c}{F(Fe II $\lambda$ 1776)}& 
\multicolumn{1}{c}{N$_{1776}$/N$_{1032}$} \\ \tableline 
Z~And&     7300$\pm$300&   24$\pm$6&      (0.6$\pm$0.3)\%   \\
V1016~Cyg& 5500$\pm$2200&  56$\pm$18&     (1.7$\pm$0.9)\%  \\
RR~Tel&    7000$\pm$1900&  10.8$\pm$3.0&  (0.3$\pm$0.1)\% \\ \tableline

\end{tabular}
\end{center}
\end{table}

\end{document}